\documentclass[prl,aps,amsmath,twocolumn,amssymb,floatfixng,showpacs,
superscriptaddress]{revtex4}

\usepackage[dvips]{graphics}
\usepackage{color}
\definecolor{dred}{rgb}{0.65,0,0}

\begin{document}

\title{\textcolor{dred}{A renormalization group study of persistent current in a 
quasiperiodic ring}}

\author{Paramita Dutta}

\affiliation{Theoretical Condensed Matter Physics Division, Saha
Institute of Nuclear Physics, Sector-I, Block-AF, Bidhannagar,
Kolkata-700 064, India}

\author{Santanu K. Maiti}

\email{santanu.maiti@isical.ac.in}

\affiliation{Physics and Applied Mathematics Unit, Indian Statistical
Institute, 203 Barrackpore Trunk Road, Kolkata-700 108, India}

\author{S. N. Karmakar}

\affiliation{Theoretical Condensed Matter Physics Division, Saha
Institute of Nuclear Physics, Sector-I, Block-AF, Bidhannagar,
Kolkata-700 064, India}

\begin{abstract}

We propose a real-space renormalization group approach for evaluating 
persistent current in a multi-channel quasiperiodic fibonacci tight-binding 
ring based on a Green's function formalism. Unlike the traditional methods, 
the present scheme provides a powerful tool for the theoretical description 
of persistent current with a very high degree of accuracy in large periodic
and quasiperiodic rings, even in the micron scale range, which emphasizes
the merit of this work.

\end{abstract}


\maketitle

\begin{figure}[ht]
{\centering \resizebox*{8cm}{2.5cm}{\includegraphics{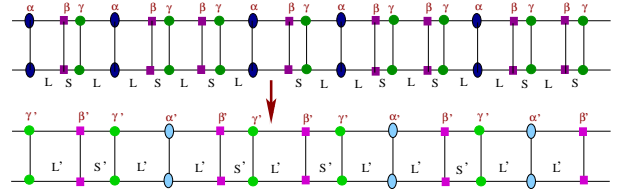}}\par}
\caption{(Color online). Upper panel: A typical realization of an $8$-th
generation fibonacci ladder composed of long ($L$) and short ($S$) bonds, 
where the sites between $L$-$L$, $L$-$S$ and $S$-$L$ bonds are labeled by 
$\alpha$, $\beta$, and $\gamma$, respectively. Lower panel: A one step 
renormalized version of the ladder.}
\label{ladder}
\end{figure}

The observation of non-decaying circular current in a metallic ring in
presence of Aharonov-Bohm (AB) flux is one of the noteworthy 
phenomena in mesoscopic physics. The measurement of low-temperature 
magnetic response of $10^7$ isolated mesoscopic copper rings to a slowly 
varying magnetic flux by Levy {\it et al.}~\cite{levy} in 1990 was the 
first experimental evidence of flux-periodic persistent current though 
the theoretical prediction was made much earlier in 1983 by B\"{u}ttiker 
{\em et al.}~\cite{butt1}. Later, many theoretical~\cite{butt2,gefen,san1,
san2,san3,san4} as well as 
experimental~\cite{mailly1,mailly2,mailly3,blu} attempts were made 
to understand the intricate role of electron-electron interaction, disorder 
and quantum interference on the phenomenon of persistent current. These
studies are mostly confined to mesoscopic systems, but interesting recent
possibilities are that micron scale DNA loops also support persistent 
current~\cite{nakamae,kim}. The DNA's are large biopolymers consisting of
thousands of atoms and exhibit both the periodic as well as quasiperiodic
arrangement of the base pairs~\cite{guo,diaz,macia1,macia2}. The 
quasiperiodic arrangements~\cite{macia3,shechtman} ({\em e.g.} fibonacci 
sequence), lacking of translational invariance, possess certain kind of 
long-range order leading to self-similar structures~\cite{snk1} and a 
number of elegant real-space renormalization group (RSRG) techniques were 
proposed to understand the unusual physical properties of quasiperiodic 
systems~\cite{kohmoto1,kohmoto2,snk2,snk3,snk4,snk5,lu}. Apart from the 
periodic ladder models, the quasiperiodic fibonacci ladder network 
(see Fig.~\ref{ladder}) has also been widely used in modeling the 
double-stranded DNA structure to study its electronic structure, charge
transfer, etc. The sensitivity of persistent current of a $1$D fibonacci 
ring to Fermi energy was investigated by some groups~\cite{jin,hu} 
calculating persistent current from the derivative of ground state energy 
of the system with respect to the applied magnetic flux. The studies on 
persistent current in the double-stranded circular DNA of higher organisms 
is highly limited~\cite{kim}, and the loop geometry of quasiperiodic 
fibonacci ladder network has yet to be investigated. A major problem for 
such biopolymers is that the system size is very large, one needs to 
diagonalize big matrices, and the numerical errors make it difficult to 
predict persistent current form the derivative of the ground state energy 
with respect to magnetic flux.

To overcome this problem, in the present communication we propose a 
RSRG technique~\cite{snk2,snk3} for the evaluation of persistent current in 
a large quasiperiodic multi-channel ring. As an illustrative example of a 
quasiperiodic multi-channel ring, here we consider a fibonacci ladder rolled
in the form of a loop. The key idea is that first we introduce the concept of 
the density of persistent current (DOPC) in terms of the retarded and advanced 
Green's functions, then using RSRG method we replace the original large loop 
by an effective loop consisting of only few effective atoms, and finally 
calculate DOPC from the Hamiltonian of the effective ring. This method is 
numerically very efficient since it involves only the iteration of few 
recursion relations of the system parameters and requires the inverse of 
small matrices in order to find the retarded and advanced Green's functions, 
and finally yields persistent current with a very high degree of accuracy.

We now describe our RSRG method for determination of persistent current in a 
quasiperiodic fibonacci ladder network in the form of a ring in which 
persistent current is induced by applying an infinitesimally slowly varying 
magnetic flux $\phi$ through the ring. This method can easily be employed to 
any periodic or quasiperiodic systems possessing self-similarity. In order 
to implement the RSRG scheme, we have to start with a general model for the 
fibonacci ladder network as illustrated in Ref.~\cite{snk2} and the ring 
shape geometry is enforced by imposing periodic boundary condition. A 
schematic diagram of a general quasiperiodic fibonacci ladder network is 
depicted in Fig.~\ref{ladder}. In the general model two types of bonds, 
namely, long ($L$) and short ($S$) bonds are arranged according to the 
quasiperiodic fibonacci sequence, and, we have three different kinds of 
atomic sites, $\alpha$, $\beta$ and $\gamma$, corresponding to the 
$L$-$L$, $L$-$S$, and $S$-$L$ vertices, respectively. Construction of 
the fibonacci sequence $\lbrace F_g \rbrace$ of generation $g$ is based on 
the rule $\lbrace F_g \rbrace$=$\lbrace F_{g-1}, F_{g-2} \rbrace$ for 
$g\ge 3$ with $F_1=S$ and $F_2=L$. Each strand of the ladder is constructed 
exactly with the same fibonacci sequence of some particular generation. 
In Fig.~\ref{ladder}, ladders corresponding to two successive generations 
of the quasiperiodic sequence are displayed where the three different 
geometrical shapes oval, square and circle represent the $\alpha$, $\beta$ 
and $\gamma$ atomic sites, respectively.

Within a tight-binding (TB) framework, the Hamiltonian of the fibonacci 
multi-channel ring threaded by a magnetic flux $\phi$ (in unit of the 
elementary flux quantum $\phi_0=hc/e$) reads,
\begin{equation}
\mbox{\boldmath$H$}=\sum_{i=1}^{N_g} \mbox{\boldmath$c$}_i^{\dag} 
\mbox{\boldmath$\varepsilon$}_i \mbox{\boldmath$c$}_i 
+\sum_{i=1}^{N_g} (e^{j \theta_{i,i+1}} \mbox{\boldmath$c$}_i^{\dag} 
\mbox{\boldmath$\tau$}_{i,i+1} \mbox{\boldmath$c$}_{i+1} + h.c.) 
\end{equation}
where, 
\begin{eqnarray}
\mbox{\boldmath$\varepsilon$}_i=\left( \begin{array}{c c}
\epsilon_i~~ v \\
v~~ \epsilon_i
\end{array} \right),
\mbox{\boldmath$\tau$}_{i,i+1}=\left( \begin{array}{c c}
t_{i,i+1}~~d_{i,i+1} \\
d_{i,i+1}~~ t_{i,i+1}
\end{array} \right),
\mbox{\boldmath$c$}_i=\left( \begin{array}{c}
c_{i,I} \\
c_{i,II}
\end{array} \right). \nonumber
\end{eqnarray}
Here, $v$ is the vertical hopping between the $i$-th sites of the two 
different strands I and II of the ladder and $N_g$ is the total number 
of sites of $g$-th generation of fibonacci lattice. The hopping integrals 
$t_{i,i+1}$ along each strand of the ladder has two values $t_L$ and $t_S$, 
corresponding to the long and short bonds respectively. Here, $d_{i,i+1}$ 
represents diagonal hopping between dissimilar sites of the two strands. 
The electron creation (annihilation) operators in the Wannier basis 
$|iI \rangle$ and $|i II \rangle$ are $c_{iI}^{\dag}$ ($c_{iI}$) and 
$c_{iII}^{\dag}$ ($c_{iII}$), respectively. The phase factor $\theta_{i,i+1}$ 
is set to $2 \pi a_{i,i+1} \phi/\Lambda$, where $a_{i,i+1}=a_L$ or $a_S$ 
represents the long or short bond lengths respectively and 
$\Lambda=\sum\limits_{i=1}^{N_g} a_{i,i+1}$ is the perimeter of the ring.

In order to implement the RSRG scheme, we first express persistent current 
in terms of the Green's function of the system. This has been achieved by 
introducing the notion of the density of persistent current $J(E)$ such that 
$J(E)dE$ gives the amount of persistent current in the energy interval 
between $E$ and $E+dE$. From the standard expression for persistent 
current~\cite{san1,san2} it can be easily be shown that DOPC is given by,
\begin{equation}
J(E) = \frac{1}{\Lambda} \mbox{Tr} \left[\sum_i a_{i,i+1} 
\left(e^{j\theta_{i,i+1}}
\mbox{\boldmath$\mathcal{G}$}_{i+1,i}-e^{-j \theta_{i,i+1}} 
\mbox{\boldmath$\mathcal{G}$}_{i,i+1} \right)\right]
\label{cur1}
\end{equation}
where, 
\begin{eqnarray}
\mbox{\boldmath$\mathcal{G}$}_{i,j}=\mbox{\boldmath$\tau$}_{j,i} 
\mbox{\boldmath$G$}_{i,j}^r
-\mbox{\boldmath$\tau$}_{j,i}^* \mbox{\boldmath$G$}_{i,j}^a.
\end{eqnarray}
$\mbox{\boldmath$G$}^r$ and $\mbox{\boldmath$G$}^a$ are the retarded and
advanced Green's functions of the system and they are defined as,
\begin{eqnarray}
\mbox{\boldmath$G$}^r(E)=\left(z^+ \mbox{\boldmath$I$}
- \mbox{\boldmath$H$} \right)^{-1}\,\mbox{and}~
\mbox{\boldmath$G$}^a(E)=\left(z^- \mbox{\boldmath$I$}
- \mbox{\boldmath$H$} \right)^{-1}
\label{green}
\end{eqnarray}
with $z^{\pm}=(E \pm j \eta)$ and $\eta \rightarrow 0^+$. It is to be 
noted that since the Hamiltonian parameters are real, initially we have 
$\mbox{\boldmath$\tau$}_{j,i}^*=\mbox{\boldmath$\tau$}_{j,i}$ but we will 
see that $\mbox{\boldmath$\tau$}_{j,i}$ becomes complex as we renormalize 
the system. Also we have $\theta_{j,i}=\theta_{i,j}$. In this expression 
we set $c=h=e=1$.

The self-similarity of the fibonacci sequence enables us to develop a RSRG 
scheme for the determination of DOPC and it ensures that a renormalized 
fibonacci lattice exactly maps to a lower generation fibonacci lattice, and 
also one can split the original fibonacci lattice into two 
fibonacci sublattices~\cite{snk2}. This allows us to express $J(E)$ of a 
given generation ring as that of a lower generation ring with renormalized 
parameters. The sum in Eq.~\ref{cur1} can be splitted into two parts 
corresponding to the two different sublattices, one comprised of only 
$\beta$ sites while the other consisting of $\alpha$ and $\gamma$ sites 
(see Fig.~\ref{ladder}), and we have,
\begin{eqnarray}
J_g(E)&= & \frac{1}{\Lambda} \mbox{Tr} \left[\sum_{i \in \alpha,\gamma} a_L 
\left(e^{j \theta_{L}} \mbox{\boldmath$\mathcal{G}$}_{i+1,i}-
e^{-j \theta_{L}} \mbox{\boldmath$\mathcal{G}$}_{i,i+1} \right)
\right. \nonumber \\
& & + \left. \sum_{i \in \beta} a_S \left(e^{j \theta_{S}} 
\mbox{\boldmath$\mathcal{G}$}_{i+1,i}-
e^{-j \theta_{S}} \mbox{\boldmath$\mathcal{G}$}_{i,i+1} \right)
\right]
\label{cur2}
\end{eqnarray}
where, the subscript `$g$' refers to the $g$-th generation of the 
quasiperiodic sequence. Now we decimate the degrees of freedom associated 
with $\beta$ sites and express $J_g$ as DOPC of a ($g-1$)-th generation 
ring with renormalized system parameters. We use the following equations 
of motion for the Green's function,
\begin{eqnarray}
\mbox{\boldmath$G$}^{r(a)}_{i+1,i}\left(z^{\pm}
\mbox{\boldmath$I$}-\mbox{\boldmath$\varepsilon$}_{\beta} 
\right)& =& e^{j \theta_L} \mbox{\boldmath$G$}^{r(a)}_{i+1,i-1} 
\mbox{\boldmath$\tau$}_L\nonumber \\
&& ~~~~+ e^{-j \theta_S} 
\mbox{\boldmath$G$}^{r(a)}_{i+1,i+1} \mbox{\boldmath$\tau$}_S^T
\label{eom1}
\end{eqnarray}
and,
\begin{eqnarray}
\left(z^{\pm}\mbox{\boldmath$I$}-\mbox{\boldmath$\varepsilon$}_{\beta}\right)
\mbox{\boldmath$G$}^{r(a)}_{i,i+1}& =& e^{-j \theta_L} 
\mbox{\boldmath$\tau$}_L^T \mbox{\boldmath$G$}^{r(a)}_{i-1,i+1}\nonumber \\
&& ~~~~+e^{j \theta_S} \mbox{\boldmath$\tau$}_S 
\mbox{\boldmath$G$}^{r(a)}_{i+1,i+1},
\label{eom2}
\end{eqnarray}
to reduce the sums of Eq.~\ref{cur2} into a single one that spans the 
sublattice comprised of only the $\alpha$ and $\gamma$ sites. We rename 
the renormalized sites as indicated in Fig.~\ref{ladder} to get the 
($g-1$)-th generation fibonacci lattice. Finally, the expression for 
$J_g(E)$ in terms of the renormalized parameters takes the form,
\begin{eqnarray}
J_{g-1} (E) &= & \frac{1}{\Lambda^{\prime}} \mbox{Tr} 
\left[\sum_{k \in \alpha^{\prime},\gamma^{\prime}} a_{L^{\prime}} 
\left(e^{j \theta_{{L}^{\prime}}} 
\mbox{\boldmath$\mathcal{G}$}_{k+1,k}^{\prime} \right.\right. \nonumber \\
& - & \left.\left. e^{-j \theta_{{L}^{\prime}}} 
\mbox{\boldmath$\mathcal{G}$}_{k,k+1}^{\prime} \right)\right. \nonumber \\
& + & \left. \sum_{i \in \beta^{\prime}} a_{S^{\prime}} 
\left(e^{j \theta_{{S}^{\prime}}} 
\mbox{\boldmath$\mathcal{G}$}_{k+1,k}^{\prime}-e^{-j \theta_{{S}^{\prime}}} 
\mbox{\boldmath$\mathcal{G}$}_{k,k+1}^{\prime} \right)\right]
\label{curprime}
\end{eqnarray}
where, $\Lambda^{\prime} (\equiv \Lambda)=\sum\limits_{k=1}^{N_{g-1}} 
a_{k,k+1}$ is the perimeter of the renormalized $(g-1)$-th generation 
fibonacci ring. The recursion relations for the parameters are,
\begin{eqnarray}
\mbox{\boldmath$\tau$}_{{L}^{\prime}}&=& \mbox{\boldmath$\tau$}_L 
\left(z^+ \mbox{\boldmath$I$} 
-\mbox{\boldmath$\varepsilon$}_{\beta}\right)^{-1} \mbox{\boldmath$\tau$}_S,
~~\mbox{\boldmath$\tau$}_{S^{\prime}}= \mbox{\boldmath$\tau$}_L, \nonumber \\
\mbox{\boldmath$\varepsilon$}_{\alpha^{\prime}}
&=&\mbox{\boldmath$\varepsilon$}_{\gamma}+
\mbox{\boldmath$\tau$}_S^T \left( z^+ \mbox{\boldmath$I$} 
-\mbox{\boldmath$\varepsilon$}_{\beta}\right)^{-1} 
\mbox{\boldmath$\tau$}_S \nonumber \\
& &~~~+
\mbox{\boldmath$\tau$}_L \left( z^+\mbox{\boldmath$I$}-
\mbox{\boldmath$\varepsilon$}_{\beta}\right)^{-1} 
\mbox{\boldmath$\tau$}_L^T, \nonumber\\ 
\mbox{\boldmath$\varepsilon$}_{\beta^{\prime}}
&=&\mbox{\boldmath$\varepsilon$}_{\gamma}
+ \mbox{\boldmath$\tau$}_S^T \left(z^+\mbox{\boldmath$I$}-
\mbox{\boldmath$\varepsilon$}_{\beta}\right)^{-1} \mbox{\boldmath$\tau$}_S, 
\nonumber\\
\mbox{\boldmath$\varepsilon$}_{\gamma^{\prime}}
&=&\mbox{\boldmath$\varepsilon$}_{\alpha}+ \mbox{\boldmath$\tau$}_L 
\left(z^+ \mbox{\boldmath$I$} 
-\mbox{\boldmath$\varepsilon$}_{\beta}\right)^{-1} 
\mbox{\boldmath$\tau$}_L^T, \nonumber \\
a_{L^{\prime}}&=&a_L+a_S,
~~a_{S^{\prime}}=a_L \nonumber\\
\theta_{L^{\prime}}&=&\theta_L+\theta_S,
~~\theta_{S^{\prime}}=\theta_L .
\label{recur}
\end{eqnarray}
Thus the calculation of DOPC for a $g$-th generation fibonacci ring reduces 
to that of a ($g-1$)-th generation fibonacci ring with renormalized 
parameters and we have the equivalence,
\begin{eqnarray}
& & J_g(\epsilon_{\alpha},\epsilon_{\beta},\epsilon_{\gamma},t_L,t_S,
a_L,a_S,\theta_L,\theta_S) \nonumber \\
&=& J_{g-1}(\epsilon_{\alpha^{\prime}},\epsilon_{\beta^{\prime}},
\epsilon_{\gamma^{\prime}},t_{L^{\prime}},t_{S^{\prime}},a_{L^{\prime}},
a_{S^{\prime}},\theta_{L^{\prime}},\theta_{S^{\prime}}).
\label{finalcur}
\end{eqnarray}
This completes one cycle of the renormalization group procedure which 
is then repeated and finally we express $J_g$ as the persistent current of 
the smallest possible fibonacci ring with renormalized parameters. The 
calculation of $J_g$ thus involves the determination of 
$N_{g_0} \times N_{g_0}$ dimensional reduced Hamiltonian just iterating the 
recursion relations (Eq.~\ref{recur}), $g_0$ being the generation index 
of the lowest possible fibonacci ring. Hence, we essentially have to 
determine the inverse of very small $N_{g_0} \times N_{g_0}$, matrices 
instead of original large $N_{g} \times N_{g}$ matrices in order to find 
the Green's functions that are needed for the evaluation of $J_g$. Knowing 
$J(E)$, persistent current at $T=0\,$K for a given chemical potential $\mu$
is obtained from the relation,
\begin{equation}
I(\phi)=\int \limits_{-\infty}^{\mu} J(E)\, dE.
\label{total}
\end{equation}
On the basis of the above theoretical formulation we now present numerical 
results for the energy spectra, DOPC and net current for the bond model of 
the fibonacci lattice for a fixed chemical potential. The bond model can be 
obtained from the general model of the fibonacci lattice by setting 
$\epsilon_{\alpha}=\epsilon_{\beta}=\epsilon_{\gamma}$, $t_L \ne t_S$ and 
$a_L \ne a_S$.

\begin{figure}[ht]
{\centering \resizebox*{7cm}{6cm}{\includegraphics{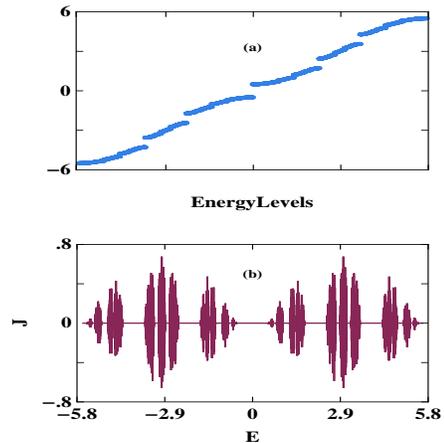}}\par}
\caption{(Color online). (a) Energy levels and (b) $J$-$E$ characteristics 
of a $17$-th generation fibonacci ring when AB flux $\phi$ is set at
$\phi_0/5$.}
\label{spectra}
\end{figure}

In Fig.~\ref{spectra}(a) we display the energy levels of a double-stranded 
fibonacci ring of $17$-th generation. The numerical values of the parameters 
are taken as, $\epsilon_{\alpha}=\epsilon_{\beta}=\epsilon_{\gamma}=0$, 
$t_L=-1$, $t_S=-\tau_n$, $v=-3$ and $d_{i,j}=0$ where $\tau_n=N_{17}/N_{16}$. 
The bond lengths $a_L$ and $a_S$ are fixed at $\tau_n$ and $1$, respectively. 
The locations of the energy eigenvalues are denoted by dots in the figure. The 
energy levels are seen to coalesce into groups in the energy spectrum. We 
have considered the vertical hopping $v$ to be high enough compared to other 
hopping integrals so that there is no overlap between the two sets of bands 
(the upper three and the lower three bands) corresponding to the two fibonacci 
strands of the ladder. The energy spectrum has the usual self-similar structure 
which was studied in detail in literature~\cite{vekilov,snk1}. One interesting 
observation is that the $J$-$E$ characteristics (Fig.~\ref{spectra}(b)) also 
exhibit the band-like structures bearing the self-similarity property 
analogous to that of the energy spectrum.
If we zoom any band in Fig.~\ref{spectra}(b), it resembles the original 
spectrum since all the energy levels contribute to the current density.
The major advantage of the RSRG approach is that it enormously reduces 
the computational load and enhances numerical accuracy in the calculation 
of persistent current of large systems like circular DNA. In the case of 
the $17$-th generation fibonacci ladder ring we have $2\times1597$, i.e., 
$3194$ sites. The evaluation of persistent current needs Green's 
functions, the determination of which requires the inversion of 
$3194\times3194$ matrices. However, our present approach based on RSRG 
arguments makes the task very simple. We can determine this current 
simply from an effective $4$-th generation fibonacci ladder ring, the 
smallest possible fibonacci ring consisting of $6$ ($=2\times3$) 
renormalized sites only. 

\begin{figure}[ht]
{\centering \resizebox*{5.5cm}{3cm}{\includegraphics{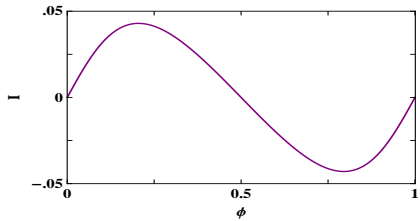}}\par}
\caption{(Color online). Current-flux characteristics for a $10$-th 
generation fibonacci ring when $v=-0.5$ and $\mu=-2.1$.}
\label{curr}
\end{figure}

It is now pertinent to raise a question whether persistent current 
calculated by this method agrees with the other existing 
methods~\cite{gefen,san1,san3,san4} or not. In order to address this 
question we have calculated persistent current using Eq.~\ref{total} 
and in Fig.~\ref{curr} we present the current-flux ($I$-$\phi$) 
characteristics for a $10$-th generation fibonacci ladder ring 
considering $v=-0.5$ and setting the other parameters values identical
to those used in Fig.~\ref{spectra}. We have 
checked that the $I$-$\phi$ curve matches very well with those obtained 
from the existing methods. To recover the on-site model for the fibonacci 
lattice from the general one, the parameters are needed to be chosen in 
the following manner: 
$\epsilon_{\alpha}=\epsilon_{\gamma} \ne \epsilon_{\beta}$, $t_L=t_S$ and 
$a_L=a_S$. Although in the present manuscript we have confined ourselves 
only to the fibonacci lattice, this method is quite general and readily 
applicable to ordered ($\epsilon_{\alpha}=\epsilon_{\gamma}
=\epsilon_{\beta}$ and $t_L=t_S$) as well as other quasiperiodic, 
{\em e.g.} copper-mean, silver-mean, etc., lattices.

To conclude, in the present communication we have introduced a method 
relying on the RSRG approach to evaluate persistent current of a 
multi-channel fibonacci ring threaded by a magnetic flux. The key idea 
is based on the concept of the DOPC such that persistent current becomes 
an integral of it over energy. Within the TB framework we have expressed 
DOPC in terms of the Green's functions of the system.
The merit of the present scheme relies on the fact that one can reduce a 
higher generation fibonacci ring to an effective smallest possible fibonacci 
ring simply iterating a set of recursion relations (Eq.~\ref{recur}) for the 
system parameters and the task of finding DOPC of the original ring 
essentially reduces to that of the effective ring. Hence, the present method 
is computationally very efficient and gives persistent current of large 
system sizes with high accuracy. This approach is quite general since it 
is applicable to any periodic or quasiperiodic system, and has great 
potential when the system size is very large. The finite temperature 
extension of this formalism is also extremely trivial, one has to multiply 
the integral in Eq.~\ref{total} just by the Fermi factor and replace the 
upper limit by infinity.

\end{document}